\documentclass[twocolumn,aps]{revtex4}
\usepackage{fleqn}
\usepackage{epsf}
\usepackage{amssymb}
\newcommand{\R}{{\mathbf{R}}}

\newcommand{\pard}[2]{{\frac{\partial {#1}}{\partial {#2}}}}
\newcommand{\deriv}[2]{{\frac{d{#1}}{d{#2}}}}

\def\ga{\mathrel{\mathchoice {\vcenter{\offinterlineskip\halign{\hfil
$\displaystyle##$\hfil\cr>\cr\sim\cr}}}
{\vcenter{\offinterlineskip\halign{\hfil$\textstyle##$\hfil\cr
>\cr\sim\cr}}}
{\vcenter{\offinterlineskip\halign{\hfil$\scriptstyle##$\hfil\cr
>\cr\sim\cr}}}
{\vcenter{\offinterlineskip\halign{\hfil$\scriptscriptstyle##$\hfil\cr
>\cr\sim\cr}}}}}
\def\vec#1{\ifmmode\mathchoice{\mbox{\boldmath$\displaystyle#1$}}
{\mbox{\boldmath$\textstyle#1$}}
{\mbox{\boldmath$\scriptstyle#1$}}
{\mbox{\boldmath$\scriptscriptstyle#1$}}\else
\hbox{\boldmath$\textstyle#1$}\fi}
\begin{document}
%_____________________________________________________________________________
\title{The influence of noise on scalings for in--out intermittency }
%_____________________________________________________________________________
\author{Peter Ashwin${}^{(1)}$, 
Eurico Covas${}^{(2)}$ 
and Reza Tavakol${}^{(2)}$
\\
}
%_____________________________________________________________________________
\affiliation{ 
1. School of Mathematical Sciences, 
University of Exeter, \\
Exeter EX4 4QE,
United Kingdom\\
2. Astronomy Unit, School of Mathematical Sciences, 
Queen Mary, University of London,\\
 Mile End
Road, London, United Kingdom
}
%_____________________________________________________________________________
\begin{abstract}
%_____________________________________________________________________________
We study the effects of noise on a recently discovered form of intermittency,
referred to as in--out intermittency. This type of intermittency, 
which reduces to on--off in systems with a skew product structure,
has been found in the dynamics of maps, ODE and PDE simulations that 
have symmetries. It shows itself in the form of trajectories that 
spend a long time near a symmetric state interspersed with short bursts 
away from symmetry. In contrast to on--off intermittency, there are 
clearly distinct mechanisms of approach towards and away from the symmetric 
state, and this needs to be taken into account in order 
to properly model the long time statistics.  We do this by using
a diffusion-type equation with delay integral boundary condition. This
model is validated by considering the statistics of a two-dimensional map with and 
without the addition of noise.
%_____________________________________________________________________________
\end{abstract}
%_____________________________________________________________________________
\maketitle
%_____________________________________________________________________________
\section{Introduction}
\label{sec_intro}
%_____________________________________________________________________________
Many dynamical systems of interest possess symmetries which force the invariance
of certain subspaces. A great deal of effort has recently gone into the study
of such systems, in particular studying the behaviour of the attractors
near their invariant subspaces on varying a parameter 
\cite{pikovsky1984,fujisakaetal1985,ashwinetal1996,plattetal1993,heagyetal1994,ottetal1994,venkataramanietal1996b,venkataramanietal1996c}.
This has included the study of systems with both normal and non-normal
parameters \cite{parameters,ashwinetal1999}.

Such systems show a variety of novel phenomena in their dynamics. In
particular, systems with normal parameters that are of skew product type 
(namely those where the transverse dynamics does not affect the dynamics
tangential to the subspace) can show on--off intermittency, which 
occurs as the result of the transversal
instability of an attractor, usually chaotic, in the invariant subspace whose
trajectories get arbitrarily close to the invariant subspace, while making
occasional large deviations away from it
\cite{ashwinetal1996,plattetal1993,heagyetal1994}. On--off intermittency can
be modelled by a biased random walk of the logarithmic distance from the
invariant subspace \cite{ashwinetal1996,plattetal1993,heagyetal1994}.

On the other hand, systems with non-normal parameters which do not have skew
product structure can show other dynamical phenomena in addition to those
present in skew product systems. These include a type of intermittency
referred to as in--out intermittency \cite{ashwinetal1999};
similar effects were noticed independently in a number of models
\cite{brooke,hasegawaetal}. Examples have been
recently found in PDE models of surface waves \cite{knobloch2000} and
in a problem of chaos control in the confinement of magnetic field lines in
toroidal fusion chambers \cite{zhangetal2000}. In the original formulation
of in--out intermittency dropping the condition that a chaotic attractor is
necessary in the invariant subspace turned out to be an important 
ingredient \cite{ashwinetal1999}, and this has subsequently been shown 
to lead to further novel phenomena \cite{laietal1999}.

\begin{figure}[!htb]
\centerline{\epsfxsize=8.80cm \epsfbox{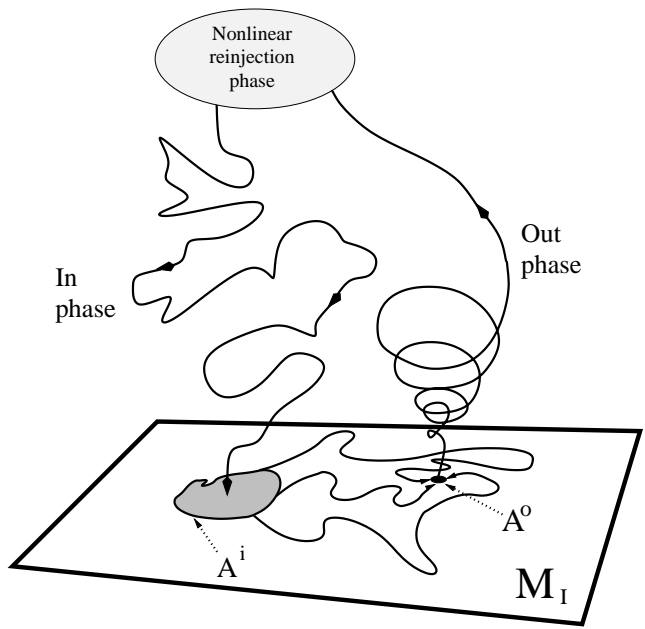}}
\caption{\label{in--out-trajectory} 
Schematic diagram showing
a typical trajectory that is in--out intermittent to an
invariant submanifold $M_I$, showing the `in' and ``out'' phases. 
The invariant submanifold $M_I$ contains an invariant set that decomposes
into a transversely attracting chaotic saddle $A^i$ and a transversely
unstable periodic orbit $A^o$ that is an attractor within $M_I$.
The `out' phase is defined by the trajectory being within the
isolating neighbourhood $U^o$ of $A^o$.
In this case, the `in' phase is modelled by a random walk 
in the logarithmic distance from $M_I$ whereas the ``out'' phase
shows uniform exponential growth away from $M_I$, shadowing the unstable
manifold of $A^o$.
}
\end{figure}

This type of intermittency is best characterized by contrasting it 
with on--off intermittency. Briefly, let $M_I$ be the invariant subspace 
and $A$ the attractor which exhibits either on--off or in--out 
intermittency.  If the intersection $A_0=A\cap M_I$ is a
minimal attractor then we have on--off intermittency, whereas (in the 
more general case) if $A_0$ is not necessarily
a minimal attractor, then we have in--out intermittency.  In the latter case
there can be different isolated invariant sets in $A_0$ associated 
with attraction and repulsion transverse to $A_0$, hence the name `in--out'. 
Another difference is that, as opposed to on--off intermittency, in the 
case of in--out
intermittency the minimal attractors in the invariant subspaces do not
necessarily need to be chaotic and hence the trajectories can instead shadow a
periodic orbit in their `out' phases \cite{ashwinetal1999}.  A schematic
representation of this scenario is depicted in Figure \ref{in--out-trajectory}.

%_____________________________________________________________________________
%_____________________________________________________________________________

There is now a good understanding of the statistical properties of
on--off intermittency
\cite{heagyetal1994,lai1996,venkataramanietal1996a,cenysetal1997b} and
some properties of in--out \cite{ashwinetal1999} intermittency where
they differ. Numerical 
support has also been obtained for both on--off
\cite{ottetal1994,hataetal1997,dingetal1997} (for experimental evidence see
\cite{hammeretal1994,yuetal1995})
and in--out \cite{cat,ctatb}.

Both of these types of intermittency rely on the presence of 
invariant subspaces. In real systems, however, invariant subspaces are
only expected to occur
approximately; either as a result of the lack of precise symmetry or due to the
presence of noise.  This has motivated a number of studies 
of the effects of noise on the statistics of on--off intermittency 
\cite{plattetal1994,cenysetal1996,ashwinetal1997,lai1997,cenysetal1997a,cenysetal1999}.
Our aim here is to make an analogous study of the effects of noise
in the case of in--out intermittency, by making a continuum version 
of the Markov model considered in \cite{ashwinetal1999} in order to
highlight the similarities and differences.  We do this 
by considering an analogue of the drift-diffusion model employed 
by \cite{ashwinetal1997,cenysetal1996,heagyetal1994,venkataramanietal1996a} 
for on--off intermittency.  

The structure of the paper is as follows. In Section \ref{sec_no_noise} 
we derive and analyze a model of in--out intermittency that consists of
a drift-diffusion equation with delay integral boundary conditions,
based on extracting the important information from a dynamical model.
We 
also discuss how to estimate parameters in the drift-diffusion model. 
Section \ref{sec_with_noise} adapts these
to include additive noise in the transverse variable as well as in the
tangential variable. We predict new transitions in the dynamics on
adding noise to the tangential variables.
Section \ref{sec_numerics} discusses the
estimation of the parameters in the model and obtains scalings
and transitions on changing the noise amplitude. These predictions 
are tested on a planar mapping given in
\cite{ashwinetal1999}. Finally Section \ref{sec_discuss} gives a discussion
and interpretation of the results.

%_____________________________________________________________________________
\section{Modelling in--out intermittency}
\label{sec_no_noise}
%_____________________________________________________________________________
\subsection{The dynamical model of in--out intermittency}
\label{sec_dyn_model}

Suppose that we have a dynamical system that evolves on $\R^n$, 
such that some subspace $M_I$ ($\dim(M_I)<n$) of $\R^n$ is dynamically 
invariant. For definiteness we consider a dynamical system
generated by iterating some smooth map $f:\R^n\rightarrow \R^n$, in which
case $f(M_I)\subseteq M_I$. If there is a minimal Milnor attractor $A$ 
for this system such that $A\cap M_I=A_0$ is {\it not} a minimal
Milnor attractor for the system restricted to $M_I$, then we say the 
attractor is in--out
intermittent \cite{ashwinetal1999}. (Recall that $A$ is a Milnor attractor
if it has a basin with positive Lebesgue measure, such that any smaller
invariant set has a basin with smaller measure. An attractor is minimal if
it contains no proper subsets that are attractors).

Suppose now that $\gamma(t)$ is a typical trajectory in the basin of $A$, such
that the $\omega$-limit set of $\gamma(t)$ is the attractor $A$. 
We assume that $A^o\subsetneq A_0$ is a Milnor attractor contained 
within $A_0$ for $f|_{M_I}$. We assume also that the only
transversely stable set in $A_0$ is some $A^i\subsetneq A_0$ which 
is a repeller for $f|_{M_I}$. Each of the invariant sets $A^{i,o}$ is
assumed to support invariant measures $\mu^i$ and $\mu^o$ that
govern the behaviour of typical trajectories in $A$ on the approach
to $A^{i,o}$.

We assume that $A^i$ is transversely attracting on average (i.e. its largest 
transverse Lyapunov exponent (L.E.) with respect to $\mu^i$ is $\lambda_T^i<0$) 
and $A^o$ is repelling on average (i.e. its largest transverse L.E.
with respect to $\mu^o$ is positive). 
We refer to $A^i$ as 
the `in' dynamics and $A^o$ as the `out' dynamics for obvious reasons;
see Figure \ref{in--out-trajectory} for a schematic representation.

\subsubsection{Identifying the `in' and `out' phases}

Suppose now that we have a projection $\Pi:\R^n\rightarrow M_I$ and a
neighbourhood $U^o\subset M_I$ containing $A^o$, such that 
$f(U^o)\supset U^o$ (i.e. it is absorbing for $f|_{M_I}$) \cite{minimalcomment}.
We identify a point $x$ in the phase space as being on `out' phase if 
$\Pi(x)\in U^o$ and as being on the `in' phase otherwise. 
Note that there is an arbitrary choice of neighbourhood
$U^o$ and projection $\Pi$; however we will be interested in
statistical properties of the `out' phases that are independent of these. 

For concreteness (and to correspond with examples studied later) we
assume that the dynamics on $A^o$ is periodic and the dynamics on
$A^i$ is chaotic (with many ergodic measures supported on $A^i$).
However, in principle the same type of model applies
as long as at least one of $A^i$ or $A^o$ is chaotic.
We are interested in modelling the asymptotic fluctuation of the
distance of some typical trajectory $\gamma(t)$ from $M_I$. 
Suppose we have a function
\begin{equation}
y(t)=\pi \gamma(t)
\end{equation}
where $\pi:\R^n\rightarrow \R$ is a smooth function such 
that $\pi^{-1}(0)= M_I$. Then we say $\pi$ {\em projects the phase 
space onto the transverse variable} 
$y$. Clearly we have $\liminf_{t\rightarrow\infty} |y(t)|=0$ but
$\limsup_{t}|y(t)|>0$. Moreover, such a transverse variable will, because of
invariance of $M_I$, spend arbitrarily long times near $y=0$; the
so-called laminar phases. The object of this paper is to give
a statistical description of the behaviour of a generic $y(t)$ measuring
the distance from $M_I$ for in--out intermittent dynamics.

\subsection{A Fokker-Planck model for in--out intermittency}

\subsubsection{In terms of a logarithmic transverse variable, $z$.}

We start with a transverse variable $y=\pi(\gamma)$
and set $z=-\ln |y|$; we model the behaviour of $z$
as follows. During the `in' phase, we model the behaviour as though
it is a linear skew product forced by the chaotic `in' dynamics
and we assume, by an appropriate scaling, that $|y|<1$
for all time. We model the behaviour as a drift-diffusion process in
$z\geq 0$ with drift $-\lambda_T^i>0$ per unit time and 
diffusion $\beta^2$ per unit time subject to reflection boundary conditions
at $z=0$. We assume that the trajectory leaks onto the 
`out' phase at a rate $\epsilon>0$ per unit time (this is given by 
the most positive tangential L.E. of $\mu_i$).

On the `out' phases we assume that there is a fixed linear expansion
forced by the periodic `out' dynamics. This translates to a
deterministic growth in the $z$ variable at a rate $-\lambda_T^o<0$.
Once $z$ reaches $0$ we assume that the trajectory is forced to
re-inject to the `in' dynamics. For convenience, from here on we define 
\begin{eqnarray*}
c&=&\lambda_T^o,\\
\lambda& =& \lambda_T^i
\end{eqnarray*}
and note that $c>0$, whereas $\lambda$ can be positive or negative in 
what follows.

Let the probability density at time $t$ of the distribution of $z$ values 
in $(z,z+dz)$ on the `in' phase be given by $P(z,t)dz$ and those on 
the `out' phase be given by $R(z,t)dz$. Our model translates to
a forward Kolmogorov equation for $P$ of the form
\begin{equation}\label{eqfpnon}
\pard{P}{t}=\pard{\Gamma}{z}-\epsilon P
\end{equation}
where the second term on the right hand side represents the leakage into the `out' phase and 
\begin{equation}
\Gamma(z,t)=\beta^2/2 \pard{P}{z}+\lambda P
\end{equation}
represents the flux of trajectories at $(z,t)$.
The dynamics for $R$ on the 
`out' phases is simply given by the hyperbolic equation
\begin{equation}
\left(\pard{}{t}- c \pard{}{z}\right) R = \epsilon P.
\end{equation}
This equation can be solved exactly to give
\begin{equation}
R(z,t)=\frac{\epsilon}{c} \int_z^\infty P \left 
(x,t-\frac{x-z}{c} \right )\,dx,
\end{equation}
which is unique up to the addition of an arbitrary function $\xi(x+ct)$.
The total probability of being in the `in' or `out' chain is then
given by
\begin{equation}
\Phi^i(t)=\int_0^\infty  P(z,t) \,dz,~~~\Phi^o(t)=\int_0^\infty R(z,t)\,dz
\end{equation}
respectively. We assume also that
\begin{equation}\label{eqfpintbc}
\Gamma(0,t)+ c R(0,t)=0
\end{equation}
at $z=0$, which corresponds to reinjecting trajectories reaching $z=0$
of the `out' chain back into the `in' chain. If we define the total 
overall probability of being in the `in' and `out' chains by
\begin{equation}
\Phi(t)=\Phi^i(t)+\Phi^o(t)
\end{equation}
then we have 
\begin{eqnarray}
\pard{\Phi}{t} 
  &=& \int_0^\infty \left ( \pard{P}{t}+\pard{R}{t} \right ) \,dz\\ \nonumber
  &=& \int_0^\infty \left (\pard{\Gamma}{z}-\epsilon P + c\pard{R}{z}+
\epsilon P \right )\,dz\\
  &=& \Gamma(0,t)+cR(0,t)\nonumber\\
  &=& 0,\nonumber
\end{eqnarray}
implying that $\Phi(t)$ is a constant. We therefore stipulate that
by normalization
\begin{equation}\label{eqfpnorm}
\Phi(t)=1
\end{equation}
for all $t$. Thus the Fokker-Planck model of in--out intermittency 
(in the absence of noise) is the closed system consisting 
of the linear equation 
(\ref{eqfpnon}) for $P(z,t)$ on $z\in[0,\infty)$ subject to the
delay integral boundary condition (\ref{eqfpintbc}) and normalization
condition (\ref{eqfpnorm}).

\subsubsection{In terms of the transverse variable, $y$.}

The drift-diffusion model in the variable $z$ can be translated into
one for the original transverse variable $y=e^{-z}$ as follows:
Let the probability density at time $t$ of the distribution of $y$ values 
in $(y,y+dy)$ on the `in' phase be given by $Q(y,t)dy$ and those on 
the `out' phase be given by $S(y,t)dy$. Note that (assuming $y>0$)
\begin{equation}
P=\left|\frac{dy}{dz}\right| Q = y Q,~~~ R =\left|\frac{dy}{dz}\right| S=y S
\end{equation}
and so the system governing $Q(y,t)$ and $S(y,t)$ is given by
\begin{eqnarray}
\label{Q-eqn}
\pard{Q}{t} & = & \frac{\beta^2}{2}\pard{}{y} \left(y\pard{}{y}(yQ) \right)
-\lambda \pard{}{y}(yQ)-\epsilon Q\\
S&=&\frac{\epsilon}{cy}\int\limits_0^y Q \left(w,t+\frac{\ln y-\ln w}{c}\right)
\label{sy}\,dw,
\end{eqnarray}
with the boundary conditions given by
\begin{eqnarray}
1 &=& \int_0^1 \left [Q(y,t)+S(y,t) \right ]\,dy \\
0 &=& \frac{\beta^2}{2}(y\pard{}{y}(yQ))-\lambda yQ +S ~~\mbox{ at }y=0
\end{eqnarray}
Observe that in terms of these variables we have
\begin{equation}
\Phi^i=\int_{0}^1 Q(y,t)\,dy,~~~~\Phi^o=\int_0^1 S(y,t)\,dy.
\end{equation}
Moreover we find
\begin{equation}
\pard{\Phi^i}{t} = \Gamma(0,t) - \epsilon \Phi^i.
\end{equation}

%_____________________________________________________________________________,
\subsection{Stationary distributions; noise free}
%_____________________________________________________________________________

Steady solutions $P(z,t)=p(z)$ of (\ref{eqfpnon}) will satisfy
\begin{equation}\label{eqstatnon}
\frac{\beta^2}{2}p_{zz}+\lambda p_z-\epsilon p=0
\end{equation}
with boundary conditions given by (\ref{eqfpintbc}) and (\ref{eqfpnorm}).
This can easily be solved to give a solution
\begin{equation}
p(z)=A e^{\mu_- z}+ A_+ e^{\mu_+ z}
\end{equation}
where
\begin{equation}
\label{eqmu}
\mu_{\pm}=\frac{-\lambda \pm \sqrt{ \lambda^2+ 2\epsilon\beta^2}}{\beta^2}.
\end{equation}
Note that if $\epsilon>0$, then
$\sqrt{\lambda^2+2\epsilon\beta^2}>|\lambda|$ and so
$\mu_-<0$ and $\mu_+>0$ always, as long as $\epsilon>0$. A similar
result was found for the Markov model of in--out intermittency
discussed in \cite{ashwinetal1999}. We therefore write
\begin{equation}
\mu=\mu_-, ~~~ A=A_-
\end{equation}
and note that the only solutions of (\ref{eqstatnon}) with finite mass 
are such that $A_+=0$. Calculating the stationary mass in the `out' 
chain we have
\begin{equation}
r(z)=Be^{\mu z}
\end{equation}
where $-c\mu B=\epsilon A$ and so $B=-\frac{\epsilon}{c \mu} A$. This 
means that
\begin{equation}
\Phi^i=\frac{A}{\mu},~~~\Phi^o=\frac{B}{\mu}.
\end{equation}
Normalizing so that the total mass is unity we have
\begin{equation}
1= \frac{A}{\mu}+ \frac{B}{\mu}
\end{equation}
which gives
\begin{equation}
A=\frac{c\mu^2}{c\mu-\epsilon}\qquad B=\frac{\epsilon\mu}{\epsilon-c\mu},
\end{equation}
thus ensuring that
the boundary condition is also satisfied
\begin{equation}
\begin{array}{l}
\Gamma(0,t)+cR(0,t)= \\
\left(\frac{\beta^2\mu^2}{2}+\lambda\mu-\epsilon\right)Ae^{\mu t}=0.
\end{array}
\end{equation}
These steady exponential distributions correspond to algebraic distributions
for the $Q(y)$ and $S(y)$. In Section \ref{secestimating} we discuss
how the free parameters in this model can be estimated from the dynamical
data.

\subsection{Contrasts with on--off intermittency}

Note that one could take a simpler dynamical model in the form of
a drift-diffusion equation but with no differentiation into
`in' and `out' phases. This is equivalent to assuming that $\epsilon=0$
in the model (i.e. reducing it to an on--off process)
and leads to an exponential probability distribution
of the form
\begin{equation}
p(z)=A e^{-\frac{\lambda}{2\beta^2} z}
\end{equation}
However, estimation of the constants $\lambda$, $\beta$ presents a problem
as we will discuss in Section~\ref{secfittoonoff}.

For in--out intermittency, the `out' phase is distinct from other 
invariant sets that we may choose within the invariant subspace in 
the following sense.
There are constants $E,F>0$ such that if any trajectory enters the 
`out' phase at a distance $d$ from the invariant subspace then
there is a minimum residence time in the `out' phase given by
\begin{equation}
E-F \ln d.
\end{equation}
In particular the minimum `out' phase residence time goes to infinity 
as $d\rightarrow 0$. 

%_____________________________________________________________________________
\section{A model for in--out intermittency with additive noise}
\label{sec_with_noise}
%_____________________________________________________________________________

The model of Section \ref{sec_no_noise} can now be easily generalized to
model in--out intermittency in the presence of unbiased additive noise.
At first we will investigate the case where the noise is added only to
the transverse variables and later examine the case of noise added
to tangential variables. It appears that noise in the transverse variables
affects scalings in a regular manner; whereas noise in the tangential
variables can lead to transitions as `in' and `out' phases merge.

\subsection{Additive noise in the transverse variables}

In presence of additive noise in the transverse variables,
we use a similar approach to that of Ashwin \& Stone \cite{ashwinetal1997} 
and Venkataramani {\em et al.} \cite{venkataramanietal1996a} to obtain 
a Fokker-Planck model of the form 
\begin{eqnarray}
\label{Q-noise-eqn}
\pard{Q}{t}&=&-\lambda \pard{(Qy)}{y}+
\frac{\beta^2}{2} \pard{}{y}\left(y\pard{(Qy)}{y}\right)+\\&&
\frac{\sigma^2}{2} \pard{{}^2Q}{y^2}-\epsilon Q,\nonumber
\end{eqnarray}
which is similar to (\ref{Q-eqn}) apart from the diffusion
term at a rate $\sigma^2/2$
corresponding to the additive noise.                               
%-------------------------------------------------
\subsubsection{Steady state with additive noise}
%-----------------------------------------------
We obtain a steady state probability density distribution in the 
in--out case, by calculating the contributions from  both `in' and 
`out' chains separately.

For the contribution from the 
`in' chain we proceed by looking at the  
steady state counterpart
of the equation (\ref{Q-noise-eqn}) obtained by demanding
$Q(y,t)=q(y)$, 
\begin{eqnarray}
-\lambda \frac{d(yq)}{dy}+
\frac{\beta^2}{2} \frac{d}{dy}\left(y\frac{d(yq)}{dy}\right)+&&\\\nonumber
\frac{\sigma^2}{2} \frac{d^2q}{dy^2}-\epsilon q=0&&
\end{eqnarray}
which can be written in the form
\begin{eqnarray}\label{steady-eqn-noise}
\frac{1}{2} \left (\beta^2y^2+\sigma^2 \right )\deriv{{}^2q}{y^2}+\left(
\frac{3}{2}\beta^2-\lambda \right)y\deriv{q}{y}&&\\\nonumber
+\left(\frac{\beta^2}{2}-\lambda-\epsilon\right)q=0.\nonumber&&
\end{eqnarray}
To solve this equation we recall that
the case with $\epsilon=0$, corresponding to on--off intermittency,
is solvable explicitly (see \cite{venkataramanietal1996a})
with the solution
\begin{equation}\label{eqQs}
q(y)=A (\beta^2 y^2+\sigma^2)^{\frac{\xi-1}{2}}
\end{equation}
where $\xi=\frac{2\lambda}{\beta^2}$.  In the case of in--out 
intermittency with $\epsilon \neq 0$, we proceed by employing 
the (singular at $\sigma=0$) change of variable
%_____________________________________________________________________________
\begin{equation}
x=-\frac{\beta^2}{\sigma^2}y^2,
\end{equation}
%_____________________________________________________________________________
to rewrite the equation (\ref{steady-eqn-noise}) in the form
\begin{eqnarray}
x(1-x) \frac{d^2 q}{d x^2}+
\left(\frac{1}{2}-x\left(1+\frac{\beta^2-\lambda}{\beta^2}\right)\right)\frac{d q}{d x}
&&\\\nonumber
-\frac{\beta^2-2\lambda-2\epsilon}{4\beta^2}q=0.&&
\end{eqnarray}
This equation can be solved in terms
of hypergeometric functions
\begin{equation}
q= A(\beta,\sigma,\lambda,\epsilon) F_{(\alpha_+,\alpha_-,1/2,x)}
\end{equation}
where
\begin{equation}
\alpha_\pm=\frac{\beta^2-\lambda\pm\sqrt{\lambda^2+2\beta^2\epsilon}}{2\beta^2} =
\frac{1+\mu_{\pm}}{2},
\end{equation}
with $\mu_{\pm}$ as before, and
\begin{equation}
F_{(a,b,c,z)}=\frac{\Gamma(c)}{\Gamma(b)\Gamma(c-b)}
\int\limits_0^1\frac{t^{b-1}(1-t)^{c-b-1}}{(1-tz)^a}dt.
\end{equation}

If $|x|\gg 1$, $x<0$ and
$Real(a-b)=2\sqrt{\lambda^2+2\beta^2\epsilon}>0$,
this solution can be approximated by using
\begin{equation}
\label{fullhyper}
\begin{array}{l}
F_{(\alpha_+,\alpha_-,1/2,x)} \\
\approx (1-x)^{-\alpha_-}\times
\frac{\Gamma(1/2)\Gamma(\sqrt{\lambda^2+2\beta^2\epsilon)}/(2\beta^2))}
{\Gamma(\alpha_+)\Gamma(1/2-\alpha_-)} \\
= K(1-x)^{-\alpha_-},
\end{array}
\end{equation}
for some constant $K$.
This then gives 
\begin{equation}\label{qs}
q(y)\approx A
\left(\sigma^2+\beta^2 y^2\right)^{-\frac{1}{2}-\frac{\mu}{2}},
\end{equation}
where $A=A (\beta,\sigma,\lambda,\epsilon) $ is the normalisation constant,
and $\mu=\mu_{-}$ as before, is valid for small $|y|$.

For the `out' chain, the steady state probability density distribution
can be obtained from (\ref{sy}) by employing the steady
state form of $Q$ in the integral in this 
expression to compute $s(y)$
\begin{equation}\label{ss}
s(y)=\frac{\epsilon}{cy} \int_0^y q(w)\,dw.
\end{equation}
%_____________________________________________________________________________
The overall steady state probability density distribution is then
given by the sum of these two contributions $q(y)+s(y)$.
%_____________________________________________________________________________
\subsubsection{Scaling of the mean first crossing time}
%_____________________________________________________________________________

To begin with, we recall that for unbiased noise on the variable $y$,
the mean is clearly
%_____________________________________________________________________________
\begin{equation}
E(y)=0.
\end{equation}
%_____________________________________________________________________________
To determine the mean crossing time through $y=0$, we require $A$ which can
be computed assuming there are no-flux boundary conditions at $y=\pm 1$
(with linear behaviour up to this point),
to give
%_____________________________________________________________________________
\begin{equation}\label{eqnormalisation}
\int_{|y|<1} P(y) \,dy=1.
\end{equation}
%_____________________________________________________________________________

Away from a blowout bifurcation point, 
it is not so easy to obtain an explicit expression for $A$
and thus for the variance
$var(y)$. However, in the low noise limit, $\sigma\ll\beta$,
we can approximate the stationary density (\ref{qs})
by the continuous function
%_____________________________________________________________________________
\begin{equation}\label{qs_approximation}
q(y) \approx \left\{
\begin{array}{ll}
A \sigma^{-1-\mu} ~~&\mbox{ for }~ |y| < \sigma/\beta,\\
A \left(\beta y\right)^{-1-\mu} ~~&\mbox{ for }~ \sigma/\beta < |y| <1,
\end{array}
\right.
\end{equation}
%_____________________________________________________________________________
and $s(y)$ is calculated from (\ref{ss}) as
\begin{equation}\label{ss_approximation}
\begin{array}{l}
s(y) \approx \frac{\epsilon A}{c}\times\\
\left\{
\begin{array}{ll}
\sigma^{-1-\mu}  ~~&\mbox{ for }~|y|<\sigma/\beta,\\
\frac{1}{-\mu}\left[(\beta y)^{-1-\mu}
-\frac{\sigma^{-\mu}(1+\mu)}{\beta y}\right]
 ~~&\mbox{ for }~|y|>\sigma/\beta.
\end{array}\right.
\end{array}
\end{equation}
In order to compare these results with those in the case of on--off, 
we proceed by computing the value of the normalisation constant $A$.
This is given by computing (\ref{eqnormalisation}) with $P(y)=q(y)+s(y)$ and gives,
for this approximation,
\begin{equation}\label{eqA}
\begin{array}{l}
A= c\mu^2\beta\left[(\epsilon-c\mu)(\beta^{-\mu}-\sigma^{-\mu})\right.
\\
~~~\left.-\sigma^{-\mu}\mu(\epsilon \ln\frac{\sigma}{\beta}(1+\mu)+(c+\epsilon)\mu))
\right]^{-1}.
\end{array}
\end{equation}
In the limit $\epsilon\rightarrow 0$ this expression reduces to the on-off
case
\begin{equation}\label{A_on--off}
A=\frac{\mu\beta}{\sigma^{-\mu}(1+\mu)-\beta^{-\mu}}
\end{equation}
equivalent to that found in \cite{ashwinetal1997}.
%_____________________________________________________________________________
Using the approximation that the stationary distributions 
$q(y)$ and $s(y)$ are approximately constant for 
$|y|<\sigma$, the instantaneous flux
from $y>0$ to $y<0$ can then be estimated as
%_____________________________________________________________________________
\begin{equation}
{\cal F}=\frac{1}{2}\sigma(q(0)+s(0))
\end{equation}
%_____________________________________________________________________________
where, as in \cite{ashwinetal1997}, we have assumed that with unbiased additive
noise approximately half of all initial points in $[0,\sigma]$ will cross over
within the next time unit. We can compute ${\cal F}$, by using the above
solutions (\ref{qs_approximation}) and (\ref{ss_approximation}) together with
(\ref{eqA}), to obtain
%_____________________________________________________________________________
\begin{equation}
\begin{array}{l}
{\cal F} = \sigma^{-\mu} \mu^2 \beta (c+\epsilon)\left[
2((\epsilon-c\mu)(\beta^{-\mu}-\sigma^{-\mu}) \right.\\
~~\left.-\sigma^{-\mu}\mu(\epsilon
\ln\frac{\sigma}{\beta}(1+\mu)+(c+\epsilon)\mu))\right],
\end{array}
\end{equation}
which in the limit $\epsilon\rightarrow 0$ reduces to the on--off formula
%_____________________________________________________________________________
\begin{equation}
{\cal F}=\frac{1}{2}\frac{\sigma^{-\mu}\beta\mu}
{\sigma^{-\mu}(1-\mu)+\beta^{-\mu}}.
\end{equation}
%_____________________________________________________________________________

This allows the calculation, in the in--out case,
of the mean first crossing time given by $M=1/{\cal F}$
analogous to \cite{ashwinetal1997}, i.e.
\begin{equation}\label{fullM}
\begin{array}{l}M = \frac{2}{\mu^2\beta(\epsilon+c)}
\left((\epsilon-c\mu)((\frac{\beta}{\sigma})^{-\mu}-1)\right. \\
\left.-\epsilon\mu \ln\frac{\sigma}{\beta}(1+\mu)+ (c+\epsilon)\mu^2)\right).
\end{array}
\end{equation}
Considering the case where $\epsilon$ is asymptotically small (i.e. small 
leakage of the `in' dynamics) we have 
\begin{equation}
\label{M_approx}
M= \frac{2}{\beta}
\left[
1+ \frac{1}{\mu} \left(\left(\frac{\beta}{\sigma}\right)^{-\mu}-1\right) +
O(\epsilon, \mu)\right].
\end{equation}
Furthermore, if we are close to marginal stability on the `in' chain,
$\lambda= 0$, we have $\mu= -\frac{\sqrt{2\epsilon}}{\beta}$ and so
in the limit $\epsilon\rightarrow 0$ we recover the expression in \cite{ashwinetal1997}, namely
%_______________________________________________________________________
\begin{equation}
M\approx\frac{2}{\beta}+\frac{2}{\beta\mu}\left(
\left(\frac{\beta}{\sigma}\right)^{-\mu}-1\right)+O(\sqrt{\epsilon}),
\end{equation}
%_______________________________________________________________________
with an order 
$\sqrt{\epsilon}$ correction. Similarly, expressions can be obtained 
for other limiting cases; we give one such scaling with the
numerical results in a later section.

We have attempted to find the scaling of the mean laminar length
with noise intensity, analogous to \cite{cenysetal1996} for on--off
intermittency. However, the need to distinguish between the dynamics
of the `in' and and `out' phases means that we cannot easily reduce the
problem to a single ordinary differential equation with the
consequence that we have so far not been able to obtain an expression
as compact as that for on--off.
However, in principle the Fokker-Planck model (\ref{Q-noise-eqn})
contains all the necessary information to compute this.

\subsection{Added noise in tangential variables}
\label{sec_noise_on_tangent}

It has been noted \cite{plattetal1994} that the addition of noise 
to tangential variables in the case of on--off intermittency has 
only a minor effect
on the dynamics. This can be understood if the attractor
within $M_I$ is stochastically stable, i.e. if the probability density
with noise limits to the probability density of the natural measure in
the case of no noise. In the case of in--out intermittency, on the other 
hand, there 
will be a threshold of noise amplitude beyond which the fine structure 
in the invariant subspace is destroyed.

To be more precise, suppose we have a dynamical scenario as described in 
Section~\ref{sec_dyn_model} and the `out' dynamics $A^o$ has a basin
such that the largest neighbourhood of $A^o$ contained in the basin
has radius $\rho>0$, and the dynamics is uniformly contracting onto $A^o$
in the tangential direction at a rate $\eta<0$. We can model the
approach to $A^o$ along its weak stable manifold by a map
$x_{n+1}=e^{\eta}x_n$ where $x$ corresponds to the distance from $A^o$. 
Perturbing this map by i.i.d. noise $\xi_n$ that is uniformly distributed in
$[-\sigma\sqrt{3},\sigma\sqrt{3}]$ (such that the variance is $\sigma$) 
we obtain an iterated function system of the
form $x_{n+1}=e^{\eta}x_n+\xi_n$. We can see that fluctuations will
drive $x_n$ to exceed $\rho>0$ if
\begin{equation}\label{eq_noise_on_tangent_thresh} 
\sigma > \rho(1-e^{\eta})/\sqrt{3}.
\end{equation}
Consequently we expect that the `out' phase and `in' phase can no longer
be distinguished once the noise has reached the order of this threshold.

%_____________________________________________________________________________
\section{Numerical results and scalings}
\label{sec_numerics}
%_____________________________________________________________________________

In order to test our model of in--out intermittency (with and without
noise) we consider a simple model mapping of the plane 
introduced in \cite{ashwinetal1999}
\begin{equation}\label{eqmap}
f(x,y)=( rx(1-x)+s xy^2, 1.82 e^{-x}y-y^3),
\end{equation}
which has two parameters $r\in[0,4]$ and $s\in\R$.
We can view this as a map
of $\R^2$ to itself that leaves $M_I=\R\times\{0\}$ invariant. 
If $s=0$, the map has the form of a skew product over
the dynamics in $x$, i.e.\ it can be written as
%_____________________________________________________________________________
\begin{equation}\label{eqskewprod}
f(x,y)=(h(x),g(x,y))
\end{equation}
%_____________________________________________________________________________
with
\begin{equation}\label{skewdef}
h(x)=rx(1-x) \ \mbox{and} \ g(x,y)=1.82 e^{-x}y-y^3,
\end{equation} where $x\in M_I$.
If we fix $r$ and vary $s$ we see that the latter two
parameters do not affect the map restricted to $M_I$ and so are 
normal parameters for the system restricted to $M_I$.
 
\begin{figure}[!htb]
\centerline{\epsfxsize=8.80cm \epsfbox{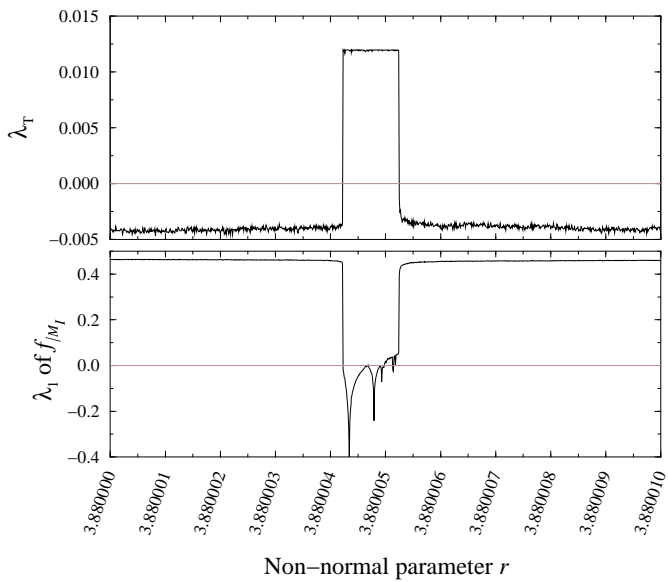}}
\caption{\label{lyapunov.exponents.in--out} 
L.E. $\lambda_{1}$ in the tangential direction and
$\lambda_T$ in the transverse direction for 
the map ({\protect \ref{eqmap}}) with
$s=-0.3$ and varying $r$. Initial conditions chosen to evolve to the
attractor in $M_I$, i.e. with $y=0$. Observe the `periodic window' coincides
with transverse instability of the attractor in $M_I$. In this region one
can find in--out intermittent attractors for the full map that are not
contained in $M_I$. Note that the attractor with transverse L.E. approximately
$-0.005$ remains as a chaotic saddle during the periodic window; this
saddle controls the in-phase.
}
\end{figure}

\begin{figure}[!htb]
\centerline{\epsfxsize=8.80cm \epsfbox{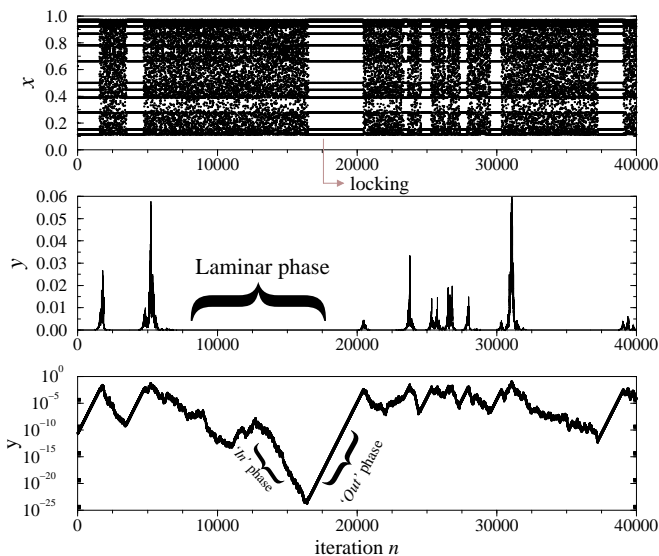}}
\caption{\label{in.out.time.series} 
Example
of time series generated by iterating (\protect{\ref{eqmap}})
from a randomly chosen initial condition, after transient effects have
died out. The bottom plot shows $y$ in
a logarithmic scale, and the top second and the middle plots show
$x$ and $y$ on linear scales.
Parameter values are fixed at $(r,s)=(3.88615,-0.3)$.
}
\end{figure}

An example of the
behaviour of the transverse and tangential L.E.s around
a window of periodicity for which the map (\ref{eqmap}) shows in--out
intermittency is depicted in Figure \ref{lyapunov.exponents.in--out}.
%_____________________________________________________________________________
%_____________________________________________________________________________
We have also shown in Figure \ref{in.out.time.series} a typical time series
corresponding to the in--out intermittent behaviour produced by this map. The
top panel clearly demonstrates 
windows of periodic locking (corresponding to the `out' phases), 
interspersed by chaotic windows (corresponding to the `in' and
`reinjection' phases). One can also clearly see from the bottom panel 
the exponential growth in the amplitude of the transverse
variable $y$ during the `out' phases.

%_____________________________________________________________________________
%_____________________________________________________________________________

To study the effects of noise on in--out intermittency, it is informative 
to compare it with the analogous studies of on--off. To do this we chose 
two sets of values for the control parameters $r$ and $s$ (namely 
$r=3.8800045$, $s=-0.3$ and $r=3.82786$, $s=0$) in the map
(\ref{eqmap}), corresponding to in--out and on--off intermittencies 
respectively. We perturbed the map with uniform noise on $[0,\zeta]$
for the $x$ dynamics and $[-\zeta,\zeta]$ for the $y$ dynamics.
We have considered the two cases above where the noise is imposed 
(I) on the tangential variable $x$ and (II) on the transverse variable $y$. 

\subsection{Estimating the parameters for the noise-free model}
\label{secestimating}

\begin{figure}[!htb]
\centerline{\epsfxsize=8.80cm \epsfbox{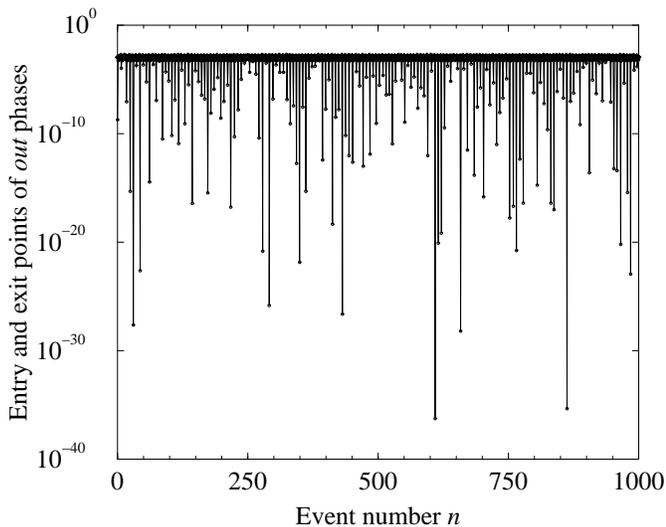}}
\caption{\label{figoutphases}
A plot of $\ln(y)$ at the entrance at the $n$th period, showing
the value at the start of the `out' phase (lower point)
and that at the exit from the `out' phase (upper point) for in--out
intermittency, plotted against event number. Observe that 
exit never occurs closer than a certain distance from 
the invariant manifold at $\ln|y|=-\infty$. The events appear to
be independent and uniformly distributed in time.
}
\end{figure}

\begin{figure}[!htb]
\centerline{\epsfxsize=8.80cm \epsfbox{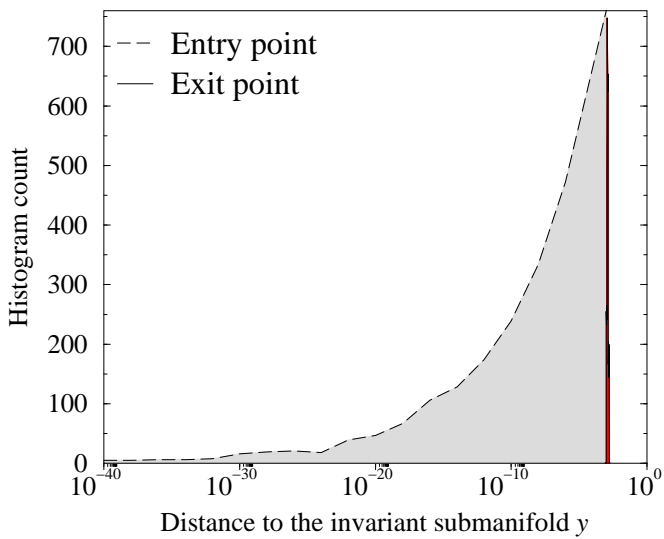}}
\caption{\label{histogram.entry.exit.points}
A histogram of the entrance to the `out' phases
and the exit from the `out' phases for in--out
intermittency 
for the trajectory considered in Figure~{\protect \ref{figoutphases}}.
}
\end{figure}

Observe that the noise-free model for in--out intermittency, after
a suitable non-dimensionalisation of the transverse variable has
four parameters; $\lambda$, $\beta$, $\epsilon$, $c$ with an arbitrary
choice for application of the boundary condition at $y=1$ after
a suitable rescaling of $y$. We estimate these parameters as follows. 
We take a trajectory such that the transient has decayed 
and we can identify parts of the trajectory as `in' or `out' phase
by choice of a suitable $U^0$.  For the map (\ref{eqmap}) with 
$r=3.8800045$ and $s=-0.3$, we can define the trajectory $(x_n,y_n)$
as being in the `out' phase if
$$
\min\{|x_{n}-p_{i}|~:i=1\ldots12\} < 10^{-4}
$$
i.e. if it approaches the period 12 attractor $\{p_{i}~:i=1\ldots12\}$ for $y=0$ to within
$10^{-4}$. Using this criterion, we have depicted in Figure \ref{figoutphases}
the values of the transverse variable $y$ at the entrance and exit of the
`out' phases identified by this procedure. Note that the exit point
is more or less constant while the entrance point is distributed in
a manner consistent with exponential distribution
in $z=-\ln |y|$ (see Figure~\ref{histogram.entry.exit.points}).
%_____________________________________________________________________________
%_____________________________________________________________________________
%_____________________________________________________________________________
%_____________________________________________________________________________

\subsubsection{Estimating $\epsilon$.}

%Regression coefficient (SLOPE)          = -0.0001798544
%Standard error of coefficient           = 4.802767e-06

\begin{figure}[!htb]
\centerline{\epsfxsize=8.80cm
\epsfbox{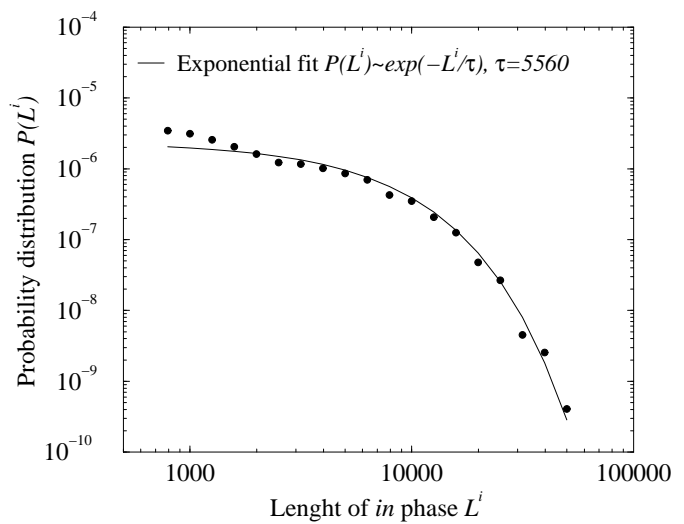}}
\caption{\label{histogram.lenght.in.phase}
The histogram of the lengths of the `in' phases 
for the trajectory considered in Figure~{\protect \ref{figoutphases}}.
This was used to calculate $\epsilon$ from the inverse of the average
length of the `in' phases $\tau$.
}
\end{figure}

Since our model implies that probability $\Phi^i$ decays from the `in' phase
at a rate $\epsilon$ per unit time, this corresponds to an exponential 
distribution of lengths of the `in' phases with average length of
`in' phase being $1/\epsilon$. Hence
\begin{equation}
\epsilon=\frac{1}{L_i},
\end{equation}
where one can easily approximate the quantity
\begin{equation}
L_i=\{\mbox{Average length of `in' phase}\}.
\end{equation}
For the map (\ref{eqmap}) with $r=3.8800045$ and $s=-0.3$,
we estimate $L_i\sim 5600\pm 200$ and so 
$\epsilon \sim  0.00018 \pm 5\times 10^{-6}$; see Figure \ref{histogram.lenght.in.phase}.

%_____________________________________________________________________________
%_____________________________________________________________________________

\subsubsection{Estimating $c$.}

This is simply the largest transverse L.E. for $A^o$,
and can in some cases be obtained analytically. 
It can also be estimated numerically as the average growth rate of $|y|$ 
during the `out' phases. For the example of in--out intermittency 
discussed above we can compute $c$ to be 
\begin{equation}
c\sim \frac{0.1434}{12}= 0.01195 \pm 10^{-5}
\end{equation}
which is the transverse L.E. of the attracting period 12 orbit in $M_I$. 

\subsubsection{Estimating $\lambda$ and $\beta$.}

These parameters can be obtained by examining the average rate of growth during
`in' phases. More precisely, we pick a threshold $z_{th}=-\ln(y_{th})$
which is large and time $T>0$, and then examine all instances
where the trajectory starts in the `in' phase at $z(t_0)=z_{th}$ and 
remains in the `in' phase for at least a time $T$.

Note that $T$ needs to
be chosen so that $z(t)$ does not get too small (i.e. $y(t)$
does not get too large) for $t_0<t<t_0+T$ and one needs to be careful to
avoid limiting the trajectory in such a way that may condition
the mean or variance we are trying to measure, for example by choosing the
threshold in $y$ to be too large, or by choosing $T$ to be so long that one will enter the
nonlinear range. 

\begin{figure}[!htb]
\centerline{\epsfxsize=8.80cm \epsfbox{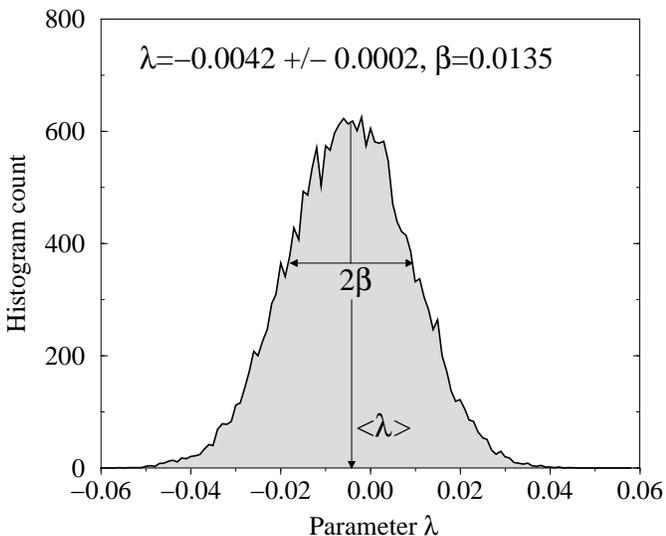}}
\caption{\label{estimate.lambda}
Histogram of the distribution of calculated values of
$\lambda$ and estimation of $\lambda$ and $\beta$ from the in phases
for the trajectory considered in Figure~{\protect \ref{figoutphases}}.
(see text for details).
}
\end{figure}

Subject to this, we can approximate $\lambda$ as the average value of
\begin{equation}
(\ln(y(t_0+T))-\ln(y(t_0)))/T
\end{equation}
over this ensemble of `in' phases.
Similarly $\beta$ can be found as the standard deviation of this quantity
from its mean value, per unit time. For the example of in--out intermittency
in map (\ref{eqmap}) discussed above, we used up to 20,000,000 points of the
trajectory, with $T=100$ and an ensemble of in-phase segments of the
same trajectory with $z_{th}\sim 30$ to find that 
\begin{equation}
\lambda\sim -0.0042,~~~\beta\sim 0.0135.
\end{equation}
where there is an expected maximum error of approximately $5\%$ (see Figure
\ref{estimate.lambda}).

%_____________________________________________________________________________
%_____________________________________________________________________________

\subsubsection{Check: an independent estimate of $\mu$.}

Recall that the ratio of the times spent
in the `in' to the `out' phases can be obtained 
from the stationary distributions in the form
\begin{equation}
R_{io}=\frac{\Phi^i}{\Phi^o}
=\frac{A}{B}=-\frac{c\mu}{\epsilon}.
\end{equation}
Using our knowledge of $\epsilon$ and $c$ we can easily obtain $\mu$
and check this against the theoretical prediction (\ref{eqmu}),
given approximations of the quantities
$
\Phi^i=$ \{ Asymptotic proportion of time spent 
on the `in' phase \}
and 
$
\Phi^o=$ \{Asymptotic proportion of time spent on  the `out' phase\} $=1-P^i$.  
For the case of in--out intermittency considered above, the above
estimates of the parameters imply that 
$$
\mu\sim -0.0451,
$$
which allow us to obtain 
a numerical estimate
for $R_{io}\sim 3.02$ with an estimated maximum error of $15\%$.

Using instead the measured ratio of average length of `in' to `out' phases we
obtain an estimate for $R_{io}= 2.56$ with an estimated maximum error of 
approximately $10\%$. These two estimates of $R_{io}$ clearly agree to within
estimated maximum error.

Note that the neighbourhood $U^0$ of the `out' dynamics $A^0$ must be
chosen such that the $U^0\cap M_I$ is forward invariant. It can be
chosen as small as desired, though if it is very small then we 
will not recognize `out' phases unless they come very close to $M_I$.

\subsection{Lack of fit to a Fokker-Planck model of on--off intermittency}
\label{secfittoonoff}

It is interesting to note that it is not possible to fit in--out intermittent 
dynamical data by an on--off model.
This is because the on--off model requires only two parameters, the
transverse L.E. $\lambda\geq 0$ and its variance per unit time $\beta^2$.
If we examine the attractor in $M_I$, we can compute a positive
L.E. ($c$ above), but the variance would be zero. Alternatively we can compute the
`in' phases as discussed above and obtain both a $\lambda$ and a $\beta^2>0$, but
in that case $\lambda<0$. Thus either choice will be invalid.

Alternatively one could compute $\lambda/\beta^2$ from the scaling of
the probability density near $y=0$, but then it is not clear how to
make a sensible choice for $\lambda$ or $\beta$ and therefore we
can determine only one 
of the parameters in the model.
%-------------------------------------------------------------
\subsection{Probability distribution for the case with noise}
%-------------------------------------------------------------

\begin{figure}[!htb]
\centerline{\epsfxsize=8.80cm 
\epsfbox{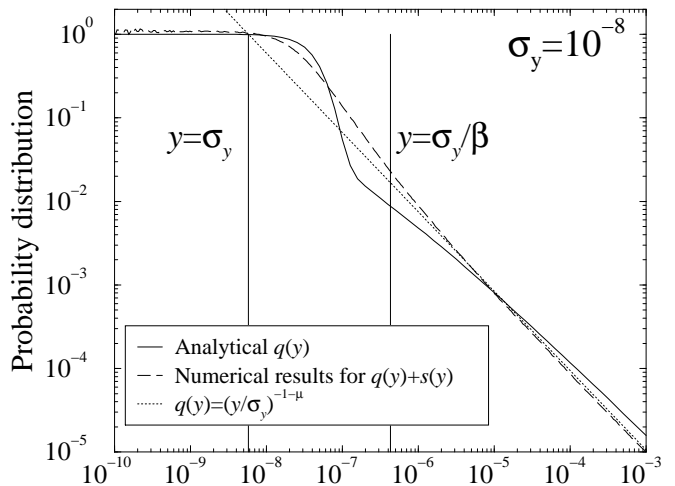}}
\caption{\label{results.q.function.y.direct.integration.PDE.Maple}
Histogram of the pdfs $q(y)$ on the in-phase and $s(y)$ on the out-phase
as a function of the transverse
variable $y$ for transverse added noise {\protect $\sigma_y=10^{-8}$}
for the examples of in--out intermittency in ({\protect \ref{eqmap}}) 
discussed in the text. Note that the analytical solution joins the
two regions, {\protect $|y|<\sigma$} and {\protect $|y|>\sigma/\beta$} 
by means of an internal boundary layer. Note that any algebraic
scaling of $q(y)\approx y^\alpha$ implies the same scaling
for $s(y)$.
}
\end{figure}

Figure \ref{results.q.function.y.direct.integration.PDE.Maple} shows
the results of three distinct ways of calculating the asymptotic
probability distribution function (pdf) of $q(y)$; resp. $s(y)$,
for a fixed noise level (in this case $\zeta_y=10^{-8}$): from the direct
integration of (\ref{steady-eqn-noise}) using the estimate values of
the parameters $\epsilon$, $c$, $\lambda$ and $\beta$; from the
full analytical solution (\ref{steady-eqn-noise}) and finally from
the direct numerical measure of the pdf of $Q(y)$.

%_____________________________________________________________________________
%_____________________________________________________________________________

\begin{figure}[!htb]
\centerline{\epsfxsize=8.80cm \epsfbox{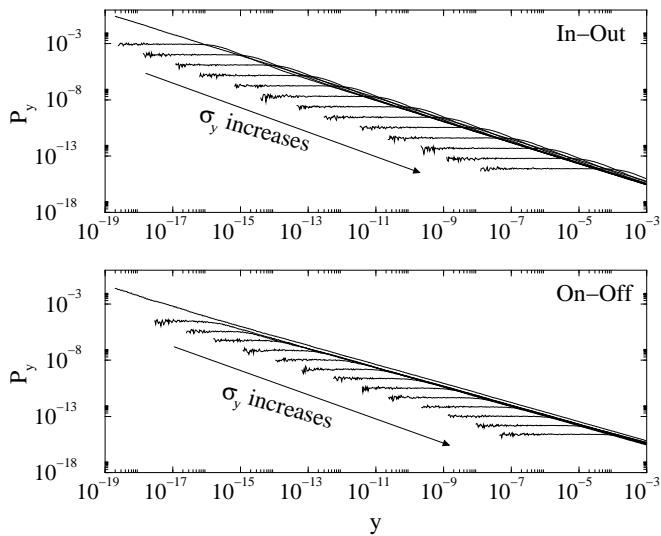}}
\caption{\label{histogram.centered}
Histogram of the pdf $Q(y)$  as a function of the transverse
variable $y$ for transverse added noise for the examples of on--off and 
in--out intermittency in ({\protect \ref{eqmap}}) discussed in the text.
Observe that there is a power law scaling across a wide range of $y$
that changes to a constant smooth density (plateau) for very small $y$. 
The turning point corresponds, as expected, to the noise level imposed; the 
response of on--off and in--out intermittency to the addition of transverse
noise can be seen to be similar.
}
\end{figure}

We have also plotted in Figure \ref{histogram.centered} the influence 
of different values of the noise level $\zeta$ on the transverse 
variable $y$ in the pdf of $Q(y)$, for both in--out and on--off cases,
and discuss the behaviour in the figure caption.

%_____________________________________________________________________________
%_____________________________________________________________________________

%_____________________________________________________________________________
\subsection{Length of the average laminar phase as a function of noise}
%_____________________________________________________________________________

\subsubsection{Noise on tangential variable}

\begin{figure}[!htb]
\centerline{\epsfxsize=8.80cm \epsfbox{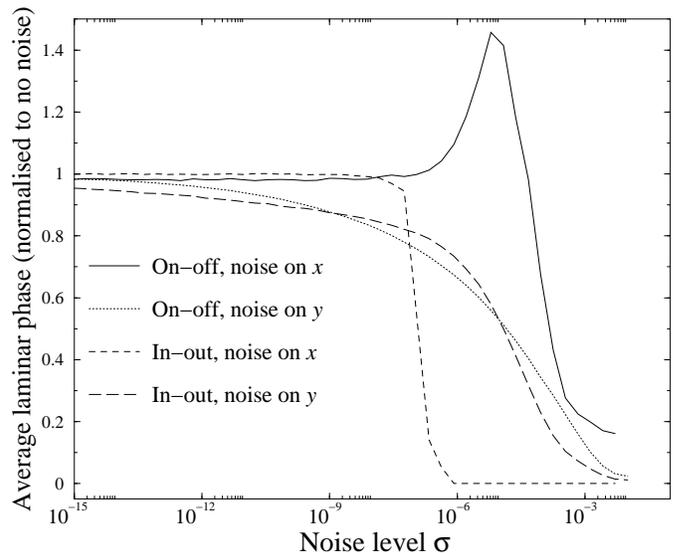}}
\caption{\label{average.laminar} 
The difference between on--off and in--out intermittency shows itself
qualitatively in the behaviour of the average laminar phase 
size as a function of the noise level $\zeta=\sqrt{3}\sigma$ for 
noise addition to the tangential variable $x$. This
illustrates the case for tangential and transverse added noise
to map ({\protect \ref{eqmap}}).
The noise on $x$ is uniformly distributed in the range 
$[0,\zeta_x]$, while that on $y$ is in the range 
$[-\zeta_y,+\zeta_y]$. The parameter values are $r=3.8800045$,
and $s=-0.3$. The decay for noise on $x$ in the in--out case corresponds
to the $x$-perturbation getting large enough to destroy the
basin of attraction of the `out' dynamics; by contrast the
addition of tangential noise to on--off intermittency only makes a difference
at much larger levels.
} 
\end{figure}

We calculated for the map (\ref{eqmap}) the scaling of the length of the
average laminar phases, 
as a function of the noise level $\zeta$.
Our results are depicted in Figure \ref{average.laminar}.
As
can be seen, in this case
there is a significant difference between the on--off and
in--out cases.  While both show very similar behaviours
for low noise levels, at higher
noise levels, the behaviours show some distinct differences.
In particular, average laminar sizes corresponding
to the in--out case drop off rather suddenly, whereas for the on--off case
there is an increase in the size of the average laminar phases before
it too drops off suddenly. This corresponds well with the discussion in
Section~\ref{sec_noise_on_tangent}, where we argued that in the case of
in--out, a sudden change in
dynamics would be expected at a certain noise level. 
For the map considered here,
we observe that the
local basin of the `out' dynamics is of the order $10^{-6}$ which
from (\ref{eq_noise_on_tangent_thresh}) would suggest a noise threshold
that is at most $10^{-6}$.

%_____________________________________________________________________________
%_____________________________________________________________________________

Also shown on this Figure are the results for the case of on--off intermittency for the same
map at a different parameter value (see caption). One would expect 
a decrease in the
average length of laminar phases until $\sigma$ is of the same order
of the threshold defining the laminar phases, which in this case is
$\sim 10^{-3}$.  This is to be contrasted with the in--out case,
where the drop off is more sudden and
occurs at much lower noise levels (i.e $\sim 10^{-6}$).
This level of noise seems to be enough to disrupt the
periodic attractor in the invariant submanifold and it is interesting
to note that the noise level at which this occurs is of the
same order of magnitude as the size of the parameter window of
periodicity.

\subsubsection{Noise on transverse variable}

\begin{figure}[!htb]
\centerline{\epsfxsize=8.80cm \epsfbox{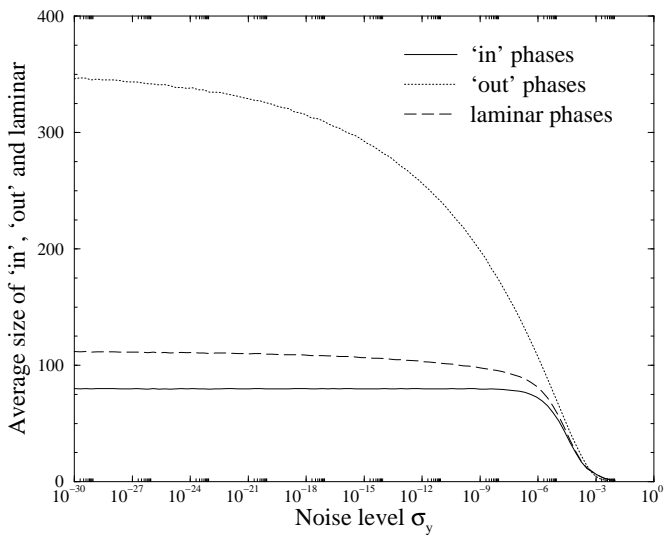}}
\caption{
\label{average.sizes.function.noise} 
Comparison of the average length of laminar phase, `in' phase and `out' phase
as a function of the noise level $\sigma$ for addition of
transverse noise in the variable $y$; parameter values 
are $r=3.8800045$ and $s=-0.3$. 
}
\end{figure}

For the noise on the transverse variable, on--off and in--out behave quite similarly in terms
of the
average laminar phases, showing a smoother decay than the case (I),
with the dramatic drop occurring around $10^{-4}$, which is closer to the
threshold level. As can be seen from
Figure \ref{average.sizes.function.noise}, the `out' chains, 
initially dominant, decay rapidly, while the `in' chains are on 
average of the same size for a wide band of noise levels $\sigma$,
up to values of about $10^{-6}$.
Note also that the actual percentage of time spent in the `in' chains 
actually increases at high noise levels before decaying to zero, 
while the time spent in the `out' chains decreases monotonically. 

%_____________________________________________________________________________
%_____________________________________________________________________________

\subsubsection{Average mean crossing time through $y=0$}

One can similarly analyze for the model the average mean crossing 
time $M$ through 
$y=0$ in presence of noise on $y$ (case (II)).  For positive
transverse L.E. ($\lambda_T>0$), that is for parameter
values away from the blowout point, one expects for the case of on--off
a typical growth given 
by
\cite{ashwinetal1997}
%_______________________________________________________________________
\begin{equation}
M\approx \frac{\beta^{1+\xi}}{\lambda}\sigma^{-\xi}+O(1).
\end{equation}
%_______________________________________________________________________

\begin{figure}[!htb]
\centerline{\epsfxsize=8.80cm \epsfbox{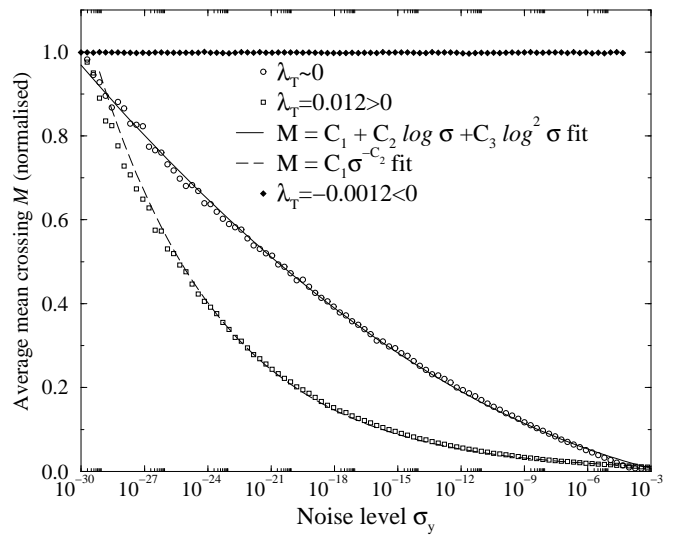}}
\caption{\label{mean.cross.function.noise.different.lyt}
Mean crossing times of variable $y$ through zero, as a function of noise level
$\sigma$, with noise added on the transverse variable
$y$, for three cases: $\lambda_T <0, \lambda_T =0, \lambda_T >0$. 
Parameter values are $r=3.8800045$, $\nu=1.82$, $a=-1$ and $s=-0.3$. The
parameter $b$ is varied to obtain normal variations on $\lambda_T$.
}
\end{figure}

For the on--off case, 
\cite{ashwinetal1997} predicts that at blowout point
the scaling has the asymptotic form ($\lambda\rightarrow 0$)
%_______________________________________________________________________
\begin{equation}
M\approx -\frac{2\ln \sigma}{\beta} +O(1).
\end{equation}
%_______________________________________________________________________
In the in--out case we find the scaling for small $|\sigma|$,
can be approximated from the expansion of (\ref{fullM})
in $\sigma$, considering $\lambda_T\ga0$, in the form
%_______________________________________________________________________
\begin{equation}
M \approx C_1 + C_2 \ln \sigma +C_3 \ln^2\sigma +O(\ln^3\sigma).
\end{equation}
%_______________________________________________________________________
We verified this for the case of in--out in the map above, by 
using the same parameter values, except for the  normal parameter $b$
which was chosen such that $\lambda_T=\ln(1.82) +b\langle x\rangle_r =0$,
in order to enable us to calculate numerically the dependence of 
the average mean crossing time $M$ through $y=0$ for
the blowout point ($\lambda_T=0$). In Figure \ref{mean.cross.function.noise.different.lyt}
we verify the above predictions of the scaling for the mean crossing time
$M$ for the three cases $\lambda_T<0$, $\lambda_T=0$ and $\lambda_T>0$.
Note in particular the case $\lambda_T>0$, which in contrast with the on--off
case needs the term in $\ln^2\sigma$ to be included (see Ashwin and Stone \cite{ashwinetal1997}
for the on--off version of Figure \ref{mean.cross.function.noise.different.lyt}).

%_____________________________________________________________________________
%_____________________________________________________________________________

\subsubsection{Probability distributions of laminar phases}

\begin{figure}[!htb]
\begin{center}
\epsfxsize=8.80cm \epsfbox{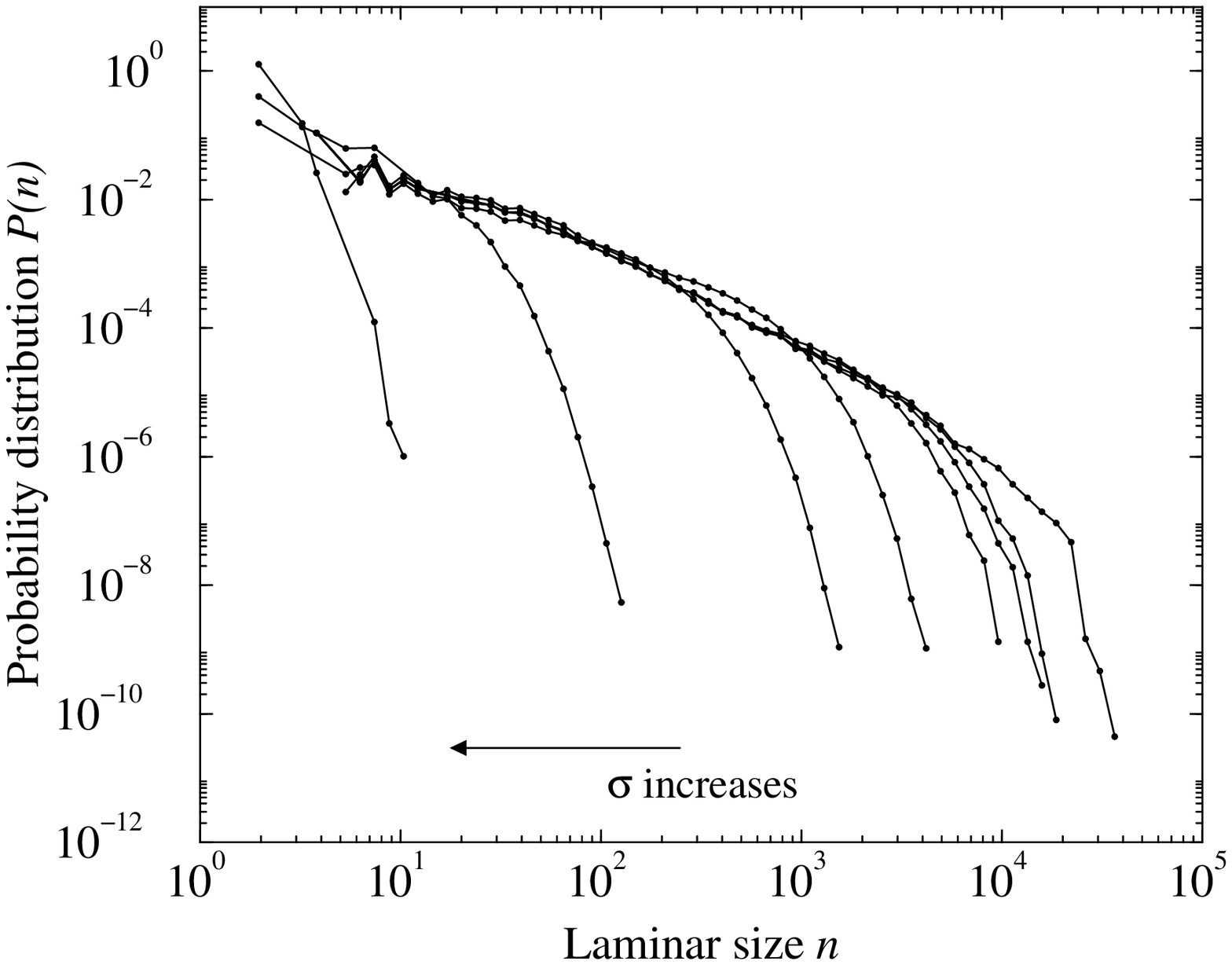}

(a)

~

\epsfxsize=8.80cm \epsfbox{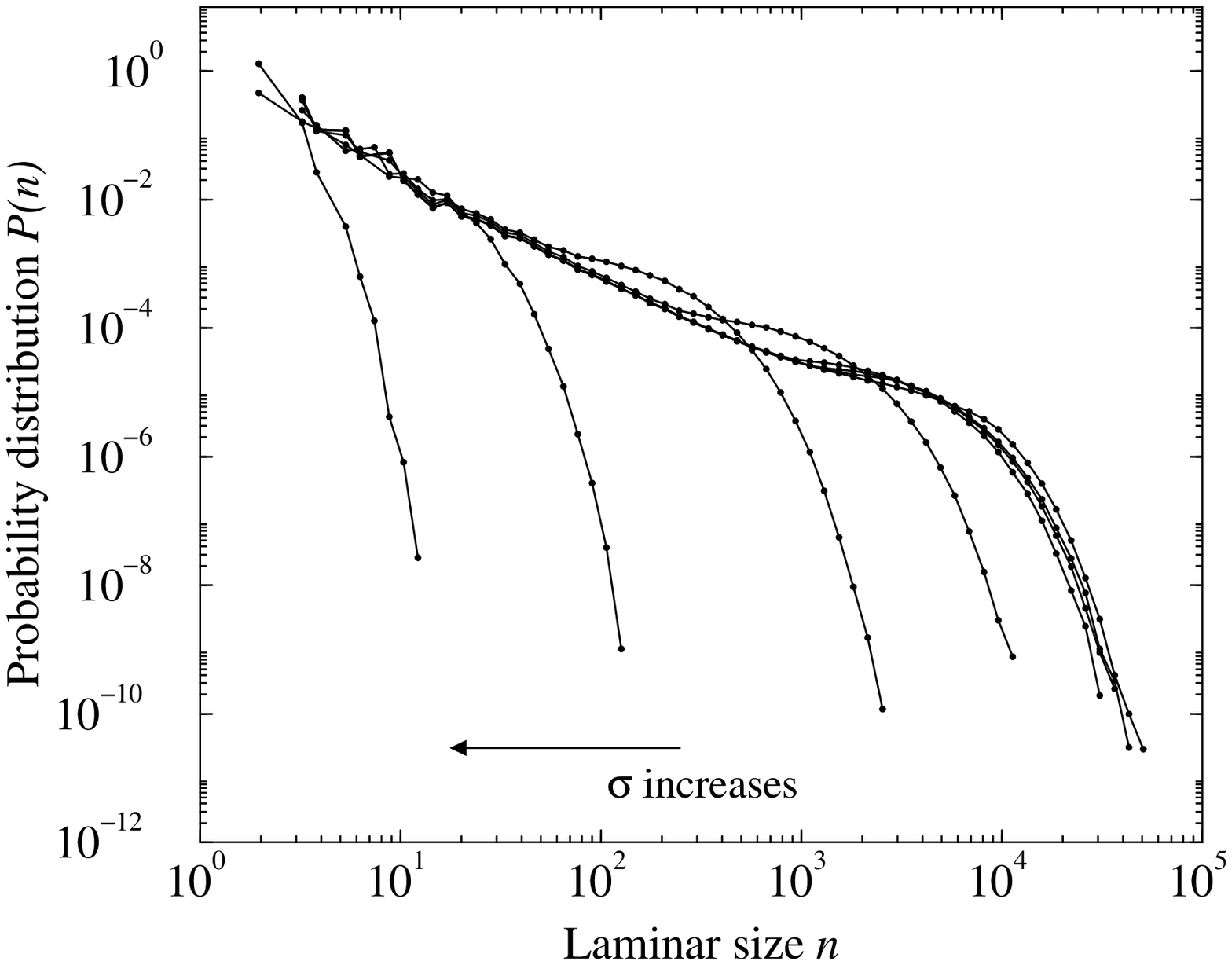}

(b)
\end{center}
\caption{\label{scalings.laminar.y.nonbiased.onoff}
Scalings of the probability distribution of laminar phases for (a) on--off
and (b) in--out intermittency with noise added in the transverse
direction. The noise perturbations on $y$ are uniform in 
$[-\sigma_y,+\sigma_y]$. Parameter values are (a) $(r,s)=(3.82786,0)$ and
(b) $(r,s)=(3.8800045,-0.3)$. Observe that the on--off statistics limit
do not show the presence of the inflexion point clearly visible in the in--out
statistics.
}
\end{figure}

We have as yet been unable to compute a closed form approximation
for the probability distribution of the laminar phases for in--out 
intermittency for the Fokker-Planck model, but
Figure~\ref{scalings.laminar.y.nonbiased.onoff} suggests that 
the scalings are analogous to those obtained in \cite{ashwinetal1999} 
for the discrete Markov model. Note the presence of an inflection point and
`shoulder' in the in--out distribution corresponding to a 
relatively high number of long laminar phases. This shoulder appears 
to persist on addition of noise. By contrast, the distribution for 
on-off laminar phases does not show such a shoulder.

%_____________________________________________________________________________
%_____________________________________________________________________________

%_____________________________________________________________________________
\section{Discussion}
\label{sec_discuss}

We have proposed a continuum model of the statistics of the transverse
variable for in--out intermittency in the form of a Fokker-Planck model
with delay integral boundary conditions to model the deterministic propagation
of probability density near the unstable manifold of the `out' phase.
This presupposes that the `out' dynamics are periodic in the invariant
manifold, but if they are not then similar models could be derived in the
form of coupled Fokker-Planck equations. Such models are then well-adapted
to model the addition of further noise.

Although in--out intermittency has a number of similarities to on--off, 
we see that there are differences in their statistical properties. 
In particular, the addition of noise `tangential' to the dynamics can 
lead, at least in our examples, to significant changes in behaviour 
at much lower noise levels than for on--off.

We have demonstrated how, given an in--out intermittent
signal, it is not possible to sensibly fit the parameters to on--off
intermittency from the dynamical data available.
There is clearly a lot more one could examine in such models, for
example, the scalings of the variance and mean first crossing times with
the various model parameters and noise level; there is work presently
in progress that aims to understand the variation of such scalings on change
of system parameters.

Implicit in our work here is the assumption that the in--out intermittent attractor 
supports a natural ergodic invariant measure, such that almost all
points attracted to the attractor will display the same stationary statistical
behaviour.
Although we do not doubt this for the models considered so far, it does
seem possible that in--out intermittency may give rise in certain
circumstances to behaviour that is not ergodic, and one needs to bear in mind
the possible existence of such behaviour.

\section*{Acknowledgements}

PA was partially supported by EPSRC grant GR/N14408 and
EC was supported by a PPARC fellowship.
%_____________________________________________________________________________

%_____________________________________________________________________________

\begin{thebibliography}{99}
\bibitem{minimalcomment}
Note
that $U^0$ cannot be open in $\R^n$, otherwise $A^o$ is an attractor for
the full systems and $A$ is not minimal.
\end{thebibliography}
\end{document}